\documentclass[prl,amsmath,aps,floats,amssymb, floatfix, superscriptaddress, twocolumn,nofootinbib]{revtex4}
\usepackage{graphicx}
\usepackage{wasysym}
\usepackage{amsfonts}
\usepackage{subfigure}
\usepackage{hyperref}
\usepackage{color}        
\usepackage{ulem}
\usepackage{lipsum}

\newcommand{\sfig}[2]{
\includegraphics[width=#2]{#1}
        }
\newcommand{\Sfig}[2]{
    \begin{figure}[thbp]
    \sfig{#1.pdf}{0.95\columnwidth}
    \caption{{\small #2}}
    \label{fig:#1}
    \end{figure}
}

\newcommand{\rf}[1]{\ref{fig:#1}}

\newcommand\be{\begin{equation}}
\newcommand\ee{\end{equation}}

\def\eea{\end{eqnarray}}
\newcommand{\ec}[1]{Eq.~(\ref{eq:#1})}

\newcommand{\eql}[1]{\label{eq:#1}}
\def\bea{\begin{eqnarray}}
\def\eea{\end{eqnarray}}
\def\vs{\nonumber\\}

\begin{document}

\title{Complementarity of Neutrinoless Double Beta Decay and Cosmology}         
\author{Scott Dodelson}        
\affiliation{Fermi National Accelerator Laboratory, Batavia, IL 60510-0500}
\affiliation{Kavli Institute for Cosmological Physics, Enrico Fermi Institute, University of Chicago, Chicago, IL 60637}
\affiliation{Department of Astronomy \& Astrophysics, University of Chicago, Chicago, IL 60637}
\author{Joseph Lykken}
\affiliation{Fermi National Accelerator Laboratory, Batavia, IL 60510-0500}

\begin{abstract}
\noindent
Neutrinoless double beta decay experiments constrain one combination of neutrino parameters, while cosmic surveys constrain another. This complementarity opens up an exciting range of possibilities. If neutrinos are Majorana particles, and the neutrino masses follow an inverted hierarchy, then the upcoming sets of both experiments will detect signals. The combined constraints will pin down not only the neutrino masses but also constrain one of the Majorana phases. If the hierarchy is normal, then a beta decay detection with the upcoming generation of experiments is unlikely, but cosmic surveys could constrain the sum of the masses to be relatively heavy, thereby producing a lower bound for the neutrinoless double beta decay rate, and therefore an argument for a next generation beta decay experiment. In this case as well, a combination of the phases will be constrained.
 \end{abstract}

\maketitle

\newcommand\mbb{m_{\beta\beta}}
\def\matm{m_{\rm atm}}
\def\msol{m_{\rm s}}

\section{Introduction}

One of the most important questions in particle physics is whether neutrinos are Dirac or Majorana particles. If they are Dirac, then their couplings to the Higgs are extremely small, thereby exacerbating the already perplexing problem of understanding mass in the Universe. If neutrinos are Majorana particles, then lepton number is violated, and there is new physics responsible for the effective operator 
$(\bar{L}\tilde{H})(\tilde{H}^TL^c)/\Lambda + h.c.$,
where $\tilde{H} = i\tau_2 H$ is the conjugated Higgs doublet, $L$ is the usual lepton doublet,
$\Lambda$ is the scale above which the new physics manifests itself, and flavor indices are suppressed.

A definitive way to resolve this question is to observe neutrinoless double beta decay (see, e.g., \cite{Bilenky:2012qi,deGouvea:2013onf}
 for reviews), which violates lepton number. 
The effective Majorana mass that governs neutrinoless double beta decay is
\bea
m_{\beta\beta} &=&  \Big\vert m_1\cos^2\theta_{12}\cos^2\theta_{13} 
\,+\, m_2 e^{2i\lambda_2} \sin^2\theta_{12}\cos^2\theta_{13}
\vs
&&\,+\, m_3 e^{2i[\lambda_3-\delta]} \sin^2\theta_{13} \Big\vert
\eql{mbb}\eea
where $(m_1,m_2,m_3)$ are the masses of the three mass eigenstates; and the mixing angles ($\theta_{12},\theta_{13}$), CP violating phases $\delta$ and Majorana phases ($\lambda_2,\lambda_3$) are the elements of the unitary matrix relating the mass and flavor eigenstates:
\begin{widetext}
\be
U = \left(
\begin{matrix}
c_{12}c_{13} & s_{12}c_{13} & s_{13}e^{-i\delta}\cr
-s_{12}c_{23}-c_{12}s_{23}s_{13}e^{i\delta} & c_{12}c_{23}-s_{12}s_{23}s_{13}e^{i\delta}
&s_{23}c_{13}\cr
s_{12}s_{23}-c_{12}c_{23}s_{13}e^{i\delta} & -c_{12}s_{23}-s_{12}c_{23}s_{13}e^{i\delta} & c_{23}c_{13}
\end{matrix}
\right)\, \left(\begin{matrix}
1 & 0 & 0\cr
0 & e^{i\lambda_2} & 0\cr
0 & 0 & e^{i\lambda_3}
\end{matrix}
\right)
\ee
\end{widetext}
where $c$ and $s$ are cos and sin. The masses are related to one another via two measured differences of mass squared: the solar mass difference $\Delta m_{12}^2\equiv m_2^2-m_1^2=\msol^2$, known to be positive and the atmospheric mass scale with
\be
 \matm^2 = \begin{cases}
  \Delta m_{23}^2 & ({\rm Normal\ Hierarchy})\cr
\Delta m_{31}^2 & ({\rm Inverted\ Hierarchy})
\end{cases}
\ee
Current and upcoming underground experiments~\cite{Barabash:2010bd,KamLANDZen:2012aa,Auger:2012gs,Ackermann:2012xja} with 10-100 kg of detector mass have the reach to explore $\mbb$ as small as 100 meV, while the ton scale experiments currently planned can push down to 10 meV. 

The experiments of the last decades have pinned down many of the neutrino parameters, so the allowed range of $\mbb$ -- and therefore the decay rate -- is becoming clearer.  In particular there emerges a key relationship \cite{Pascoli:2005zb} between $m_{\beta\beta}$ and the sum of the neutrino masses. Here we point out that cosmic surveys, which are sensitive to the sum of the neutrino masses~\cite{Abazajian:2011dt,Abazajian:2013oma}, can further narrow the allowed range of $\mbb$ and, in the future, the two sets of experiments can work together to measure one of the Majorana phases~\cite{Minakata:2014jba}. The cosmic surveys exploit the fact that the ratio of the energy density of the cosmic neutrinos to the matter density is $f_\nu=0.008 \, \sum m_\nu/{\rm 100 \, meV}$. Even this small of a fraction disrupts the delicate balance between the gravitational accretion of cold dark matter and the expansion of the universe that would otherwise produce constant gravitational potentials. A small fraction of non-clumping matter (and neutrinos are traveling fast enough not to clump on most scales) leads to decaying gravitational potentials. For example, the power spectrum of the potential is reduced by 5\% if $\sum m_\nu=100$ meV. This suppression can be inferred by measuring the cosmic microwave background temperature~\cite{Hu:2001tn} and polarization~\cite{Hu:2001kj} on small angular scales. Following the initial detections by the Atacama Cosmology Telescope~\cite{Das:2011ak} and the South Pole Telescope~\cite{vanEngelen:2012va}, the Planck satellite has now mapped the potential with 27-sigma~\cite{Ade:2013tyw} significance. Prospects for measuring the spectrum of the potential with upcoming small scale CMB polarization experiments and with galaxy surveys~\cite{Font-Ribera:2013rwa} lead to projections that $\sum m_\nu$ can be constrained at 16 meV level.

\section{Inverted Hierarchy}

If the mass hierarchy is determined by other experiments to be inverted so that $m_3<m_1,m_2$, then both $m_1$ and $m_2$ are of order $\matm$ or larger. Further, the smallness of $\sin^2\theta_{13}$ means that the last term in \ec{mbb} can be neglected so that
\bea
\mbb^{\rm inv} &\simeq &c_{13}^2\, \Big[ (m_1c_{12}^2)^2 + (m_2s_{12}^2)^2 \vs
&& + 2\cos(2\lambda_2)(m_1c_{12}^2)(m_2s_{12}^2)
\Big]^{1/2}.\eql{minv}
\eea
Fig.~\rf{numass_invdbd} shows the region allowed by current measurements for $\mbb$ and 
\be
S\equiv \sum m_\nu
\ee 
and a projected future measurement centered on a randomly chosen ``truth'' value. The width of the gray band is determined by the Majorana phase $\lambda_2$. If nature has chosen the point in parameter space indicated by the star, then the combination of neutrinoless double beta decay and cosmic surveys will narrow the allowed range; i.e., it will pin down not only the sum of the masses and the Majorana nature of the neutrino but also constrain $\lambda_2$.

 \Sfig{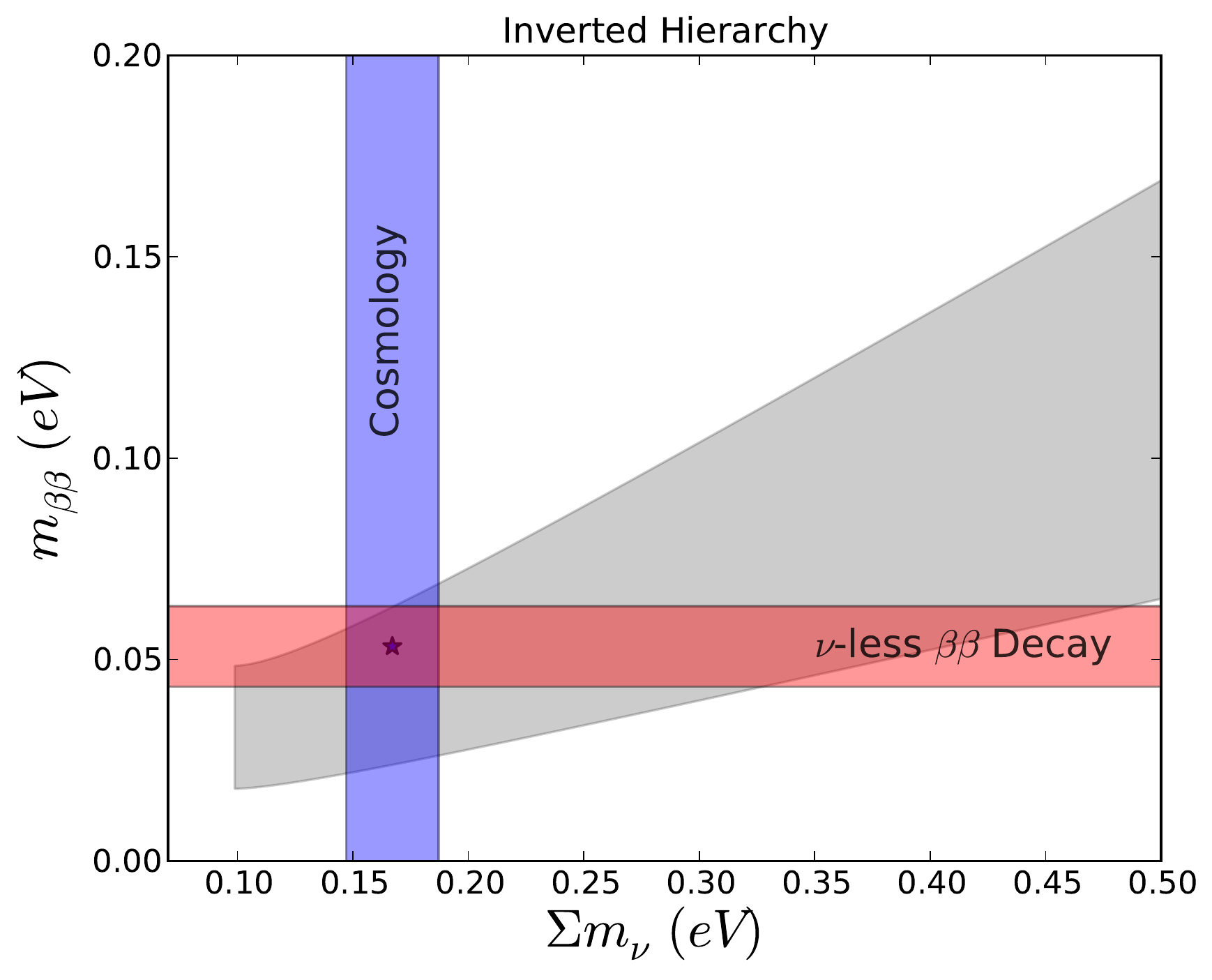}{Projected constraints on neutrino parameters from upcoming cosmic surveys (vertical), neutrino-less double beta decay experiments (horizontal), and all other current measurements (gray) assuming an inverted mass hierarchy and Majorana neutrinos.}

%

To project the error on the phase, we need to transform the projected constraints on the two parameters $p_1=\mbb$ and $p_2=S$ to a different parameter set, $(q_1=\cos(2\lambda_2),q_2=S)$. The constraints on $\vec p$ are uncorrelated, so the Fisher matrix that describes these constraints is trivial:
\be
F_{ij} = \left(\begin{matrix}
1/\Delta^2_{\beta} & 0 \cr 0 & 1/\Delta^2_S
\end{matrix}\right)
\ee
where we take $\Delta_{\beta}=10$ meV and $\Delta_S=20$ meV. Here $\Delta_\beta$ represents a combination of the uncertainty from the nuclear matrix elements \cite{Aalseth:2004hb} in the extraction of $m_{\beta\beta}$, 
together with the experimental uncertainties. For simplicity we will neglect uncertainties 
on the mixing angles, which in any case will be known with considerable precision 
from future planned experiments.
Relating the Fisher matrix of the new parameter set $(q_1=\cos(2\lambda_2),q_2=S)$ to the Fisher matrix of the old parameter set requires the transformation
\be
\tilde F_{ab} = \frac{\partial p_i}{\partial q_a}\,\frac{\partial p_j}{\partial q_b}\,F_{ij}
.\eql{fish}\ee 

Two partial derivatives of $\mbb$ are needed in \ec{fish}. The first, $\partial\mbb/\partial\cos(2\lambda_2)$. is easily obtained by differentiating \ec{minv}. The derivative with respect to $S$ is trickier but can be computed by recognizing that
\be
S = m_3 + \sqrt{m_3^2+\matm^2} + \sqrt{m_3^2++\matm^2+\msol^2}.
\ee
Differentiating both sides with respect to $S$ leads to an expression for $\partial m_3/\partial S$ at fixed $\lambda_2$. From this, $\partial m_2/\partial S = (m_3/m_2)\,\partial m_3/\partial S$ and similarly with $m_1$. Therefore, the derivative of $\mbb$ with respect to $S$ (at fixed phase $\lambda_2$) is 
\bea
\frac{\partial \mbb}{\partial S} &=& \frac{m_1m_2m_3}{\mbb\,\left[m_1m_2+m_1m_3+m_2m_3\right]}\,
\vs&\times&\Bigg[ c_{12}^4c_{13}^4 + s_{12}^4c_{13}^4 \vs
&&+ \frac{m_1^2+m_2^2}{m_1m_2} c_{12}^2c_{13}^4s_{12}^2\cos(2\lambda_2)
\Bigg].
\eea
Since $\partial \mbb/\partial S$ and $\partial \mbb/\partial \cos(2\lambda_2)$ are both non-zero, the diagonal $F$ is transformed into an off-diagonal $\tilde F$. The projected error on one parameter -- say $\cos(2\lambda_2)$ -- must be obtained by marginalizing over all possible values of the other. The simple way to do this is to compute $\tilde F^{-1}$; the diagonal components of $\tilde F^{-1}$ are the projected squared errors on the two parameters.
A simple check is that the marginalized error on $S$ -- the square root of $(\tilde F^{-1})_{22}$ -- remains the same, equal to $\Delta_S$. The error on the phase is
\bea
&&\hspace{-30pt}
(\Delta\cos(2\lambda_2))^2 = (\tilde F^{-1})_{11}
\vs
&&\hspace{-15pt}
=
\left(\frac{\partial\mbb}{\partial\cos(2\lambda_2)} \right)^{-2}\, \Big[
\Delta_\beta^2 + \left( \frac{\partial\mbb}{\partial S}\right)^2 \Delta_S^2
\Big].
\eea
Fig.~\rf{phase} shows this error as a function of $S$ for two different values of $\lambda_2$. Note that, for $S$ small and $\cos(2\lambda_2)=-1$, the projected 1-sigma error on the cosine is $0.35$, close to 6-sigma away from the $\lambda_2=0$ value.

\Sfig{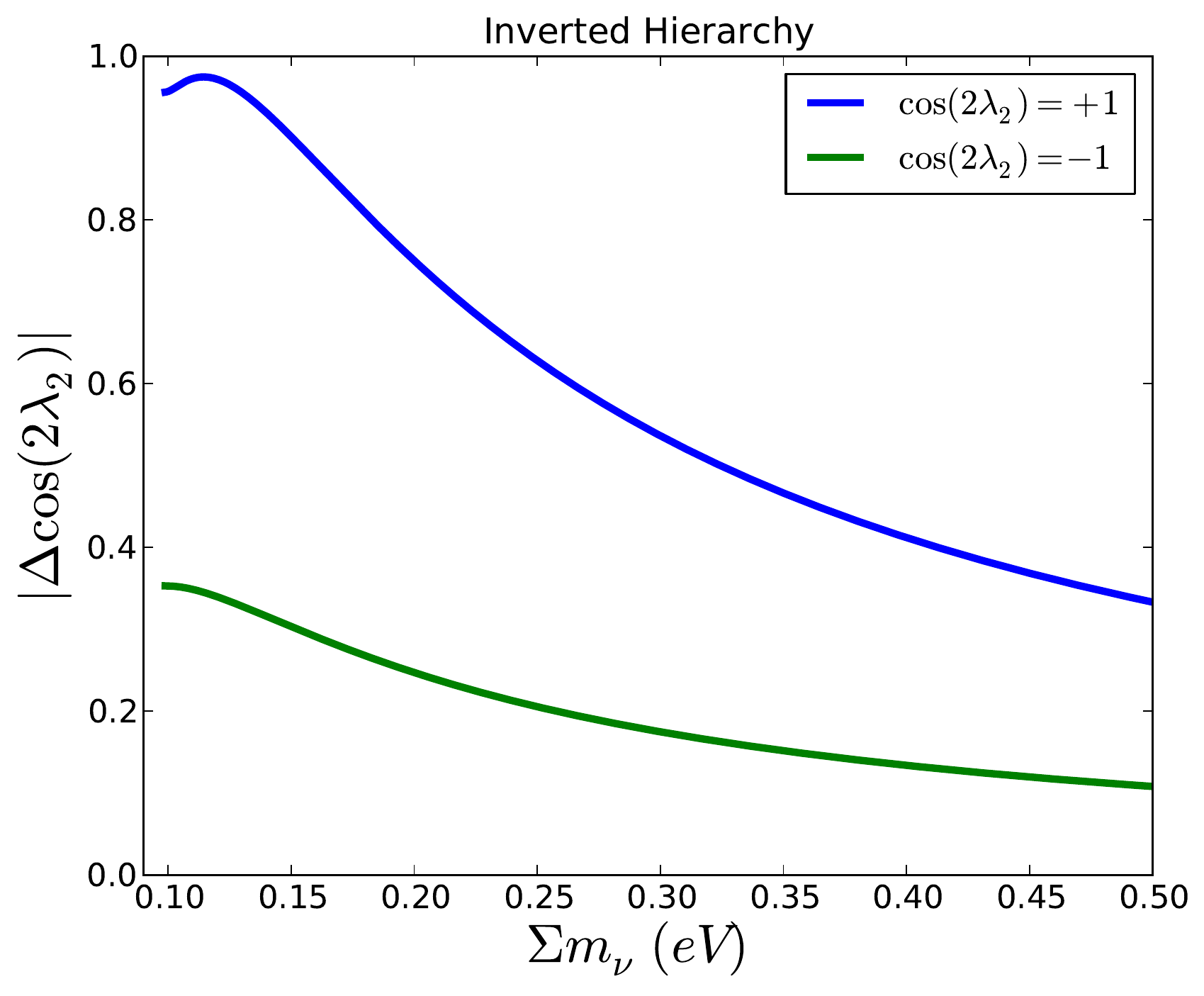}{Projected one-sigma constraint on the cosine of the Majorana phase from combined cosmic survey and neutrino-less double beta decay experiments. These constraints are relevant if the mass hierarchy is determined to be inverted.}

\section{Normal Hierarchy}

If the mass hierarchy is {\it normal} so that $m_1<m_2<m_3$, there is no guarantee that, even if neutrinos are Majorana particles, the most aggressive double beta decay experiment will see events. The parameter that determines the decay rate, $m_{\beta\beta}$, can vanish if the unknown phases conspire to make us unlucky. This is captured by the gray band in Fig.~\rf{numass_nor}, which shows that $m_{\beta\beta}$ can be arbitrarily small. However, there is an interesting synergy between the cosmological constraints and double beta decay. If the cosmological constraints point to a large value of $\sum m_\nu$, for example at the star in the figure, then we will be handed a lower limit on $m_{\beta\beta}$. The lower limit on $\mbb$ is shown as a function of $S$ in Fig.~\rf{mbbmin}.

\Sfig{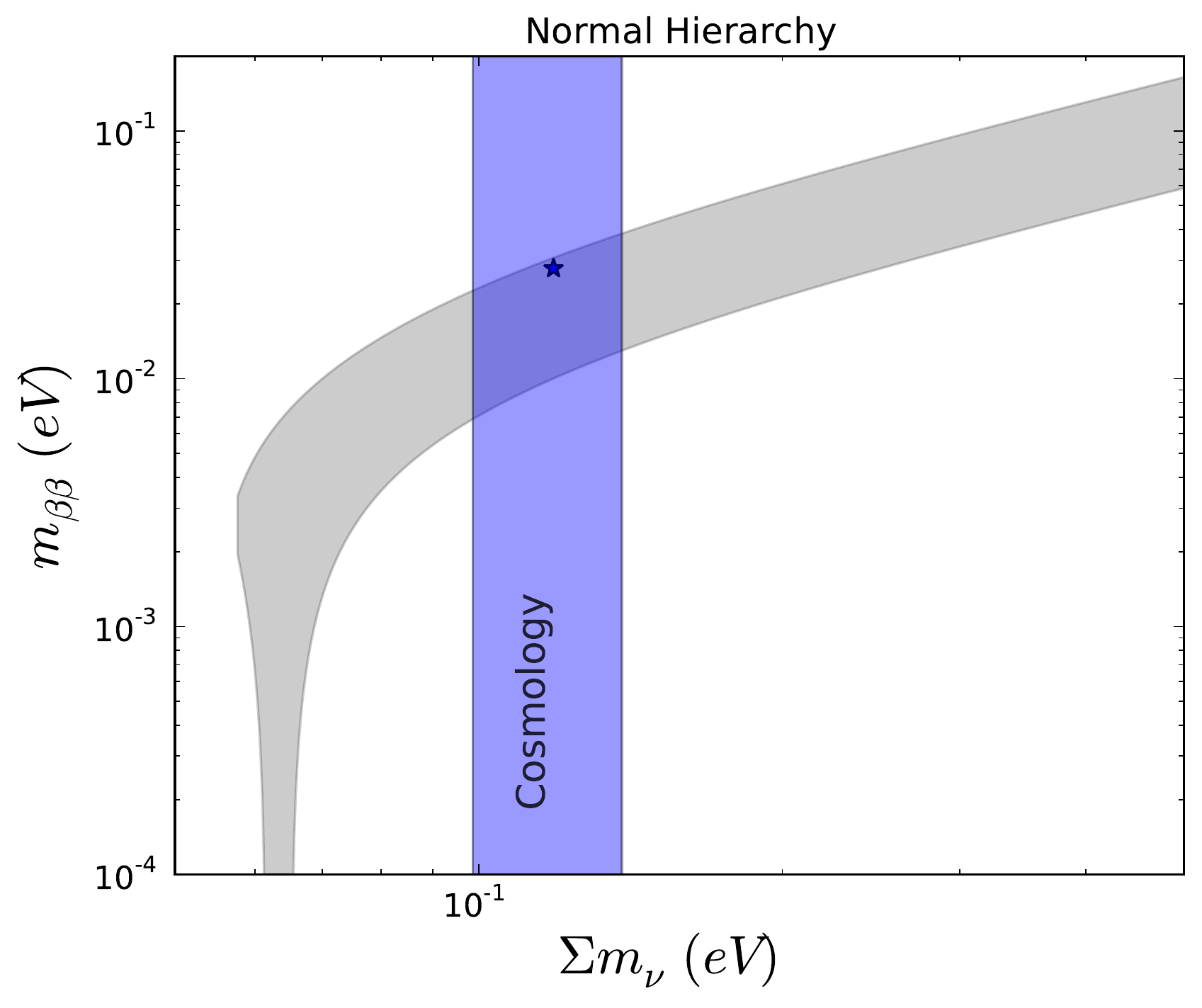}{If the mass hierarchy is normal but the sum of the masses is still relatively large, for example at the value indicated by the star, then there will be a lower limit on $m_{\beta\beta}$, a target for ambitious future double beta decay experiments.}

Therefore, upcoming cosmic surveys have the potential to motivate further neutrinoless double beta decay experiments, as we may be able to infer a lower limit on $\mbb$. In the absence of this lower limit, we will never be guaranteed an answer to the question of whether neutrinos are Majorana or Dirac particles.

\Sfig{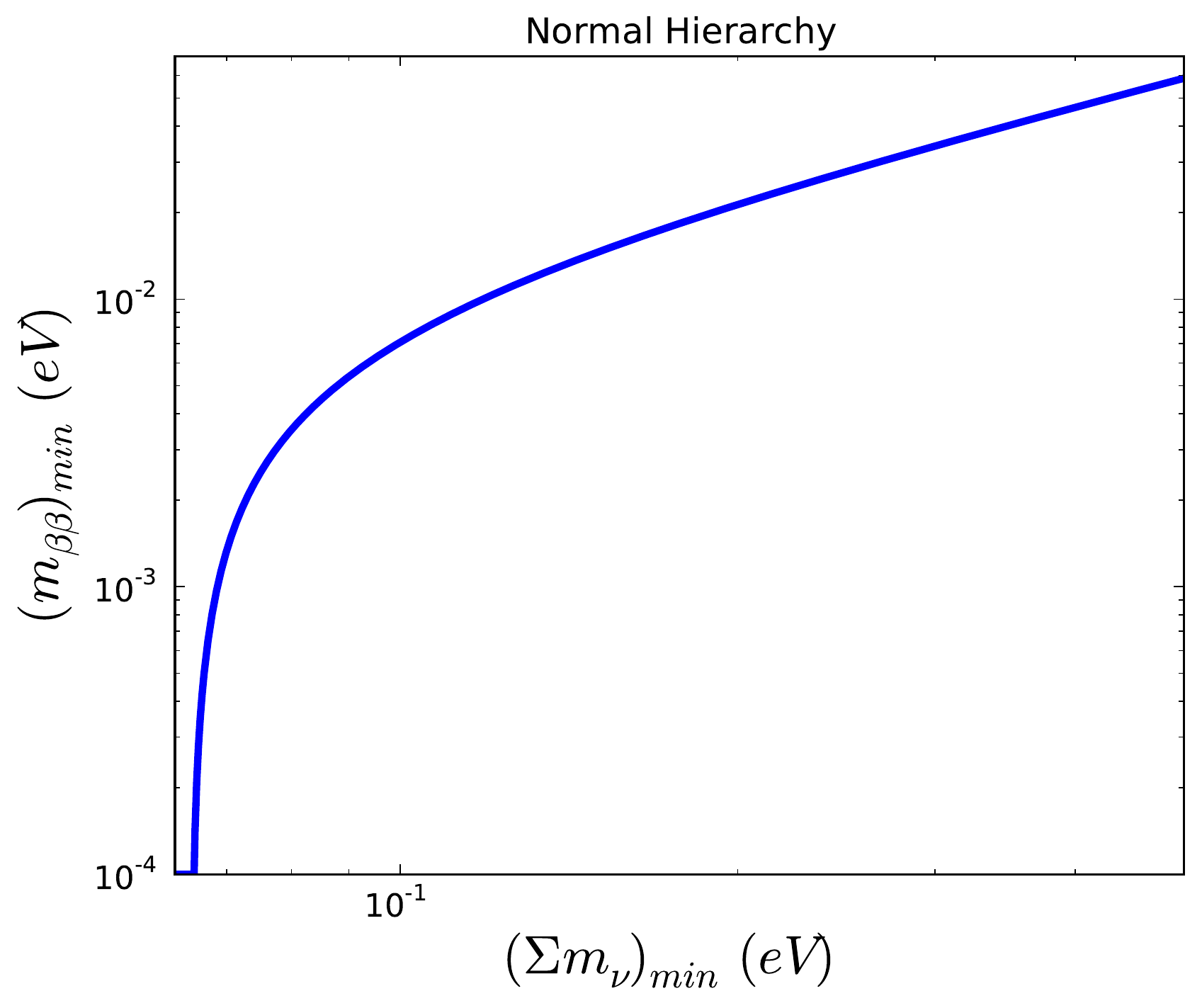}{In the normal hierarchy, the minimum value of $m_{\beta\beta}$ as a function of the lower limit on the sum of the masses that would be obtained in cosmic surveys. If the surveys find $\sum m_\nu$ is greater than $(m_\nu)_{\rm min}$, then $m_{\beta\beta}$ must be above the curve.}

{\it Acknowledgments} We are grateful to Bob Tschirhart, Andr\'e de Gouv\^ea, and Boris Kayser for useful suggestions and conversations. This work is supported by the U.S. Department of Energy, including grant DE-FG02-95ER40896. {\bf Note added}: after this work was finished, the preprint \cite{Minakata:2014jba} appeared, which has overlap with our results.

\bibliography{neutrino}
\end{document}